\documentclass[sn-mathphys,Numbered]{sn-jnl}

\usepackage{graphicx}%
\usepackage{multirow}%
\usepackage{amsmath,amssymb,amsfonts}%
\usepackage{amsthm}%
\usepackage{mathrsfs}%
\usepackage[title]{appendix}%
\usepackage{xcolor}%
\usepackage{textcomp}%
\usepackage{manyfoot}%
\usepackage{booktabs}%
\usepackage{algorithm}%
\usepackage{algorithmicx}%
\usepackage{algpseudocode}%
\usepackage{listings}%
\usepackage{pifont}
\usepackage{subfigure}
\usepackage{hhline}
\usepackage{tabularx}
\usepackage{ulem}

\theoremstyle{thmstyleone}%
\theoremstyle{thmstyletwo}%

\theoremstyle{thmstylethree}%

\begin{document}

\title[Article Title]{Preliminary Design of Detector Assembly for DIXE}


\author[1]{ \sur{Jiejia Liu}}\email{liujj21@mails.tsinghua.edu.cn}

\author[1]{ \sur{Sifan Wang}}\email{wsf18@mails.tsinghua.edu.cn}

\author*[1]{ \sur{Hai Jin}}\email{jinhai@tsinghua.edu.cn}

\author[1]{\sur{Qian Wang}}\email{qianwang0304@mail.tsinghua.edu.cn}

\author[1]{\sur{Wei Cui}}\email{cui@tsinghua.edu.cn}

\affil[1]{\orgdiv{Department of Astronomy}, \orgname{Tsinghua University}, \orgaddress{\postcode{100084}, \state{Beijing}, \country{China}}}


\abstract{Diffuse X-ray Explorer (DIXE) is a proposed X-ray spectroscopic survey experiment for the China Space Station. Its detector assembly (DA) contains the transition edge sensor (TES) microcalorimeter and readout electronics based on the superconducting quantum interference device (SQUID) on the cold stage. The cold stage is thermally connected to the ADR stage, and a Kevlar suspension is used to stabilize and isolate it from the 4 K environment. TES and SQUID are both sensitive to the magnetic field, so a hybrid shielding structure consisting of an outer Cryoperm shield and an inner niobium shield is used to attenuate the magnetic field.  In addition, IR/optical/UV photons can produce shot noise and thus degrade the energy resolution of the TES microcalorimeter. A blocking filter assembly is designed to minimize the effects. In it, five filters are mounted at different temperature stages, reducing the probability of IR/optical/UV photons reaching the detector through multiple reflections between filters and absorption. 
This paper will describe the preliminary design of the detector assembly and its optimization.
}


\keywords{X-ray Astronomy, DIXE, magnetic shielding, blocking filter}



\maketitle

\section{Introduction}\label{sec1}

Diffuse X-ray Explorer (DIXE) is an X-ray spectroscopic survey experiment proposed for the China Space Station (CSS) \cite{Jin_DIXE_LTD}. DIXE is going to conduct a three-year high-resolution X-ray spectroscopic all-sky survey, probing the physical properties of the hot baryons in the Milky Way, and tracing the Galactic feedback processes. The primary scientific objective of DIXE is to gain insights into the origins of large-scale structures of hot gas in the Milky Way, including the Local Bubble, the eROSITA Bubble, Cygnus Superbubble (and other superbubbles), as well as the Galactic halo \cite{Jin_DIXE_LTD}. 

\begin{figure}[H]
    \centering
    \includegraphics[width=\textwidth]{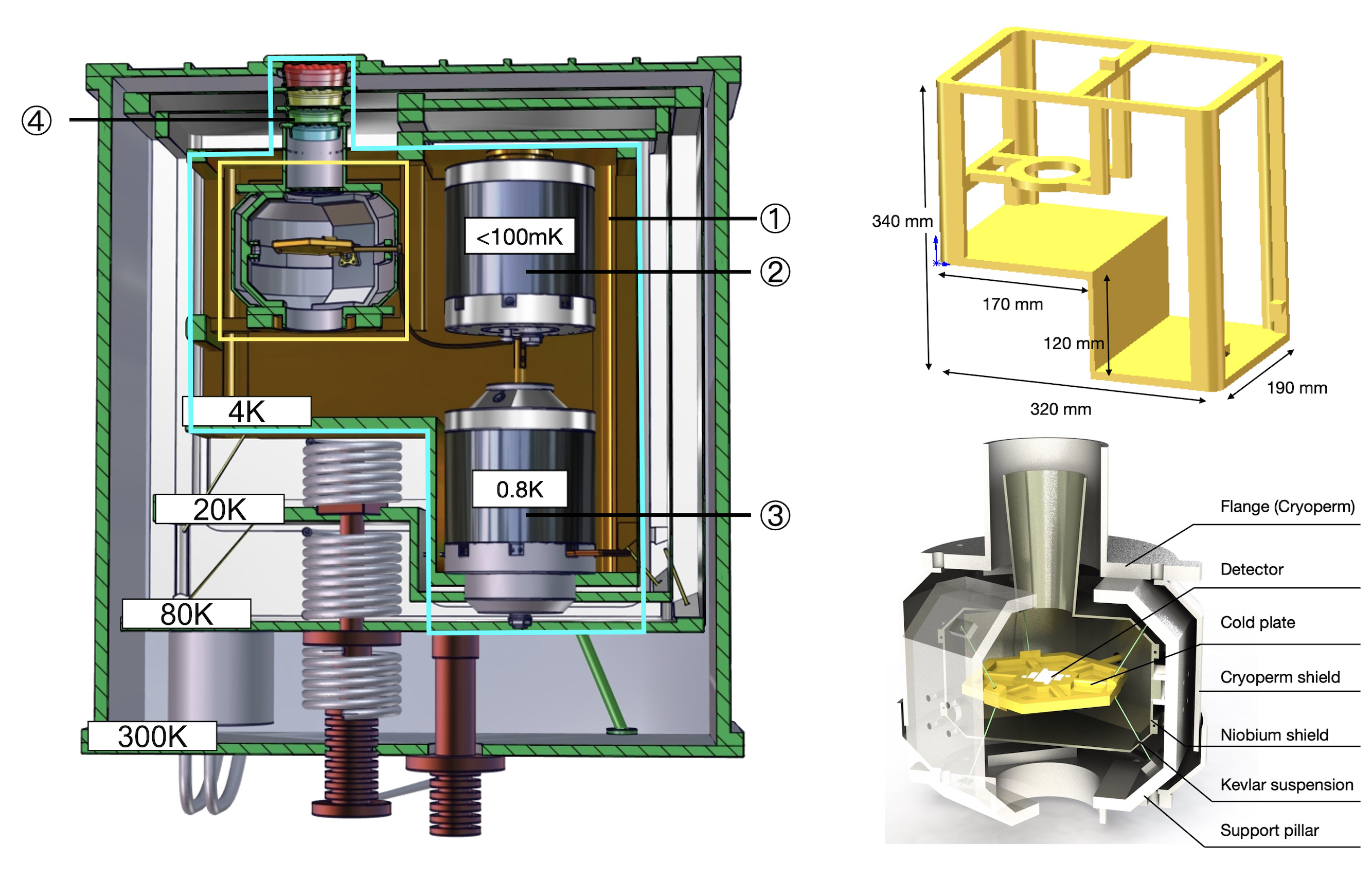}
    \caption{The preliminary design of the detector assembly. The \textit{left} panel shows the internal layout of the payload. The detector assembly (DA, highlighted in yellow) and the ADR cooling system are enclosed in the cyan box. \ding{172}: A supporting frame is affixed to the 4~K stage of the cooling apparatus. \ding{173}: The FAA stage of the adiabatic demagnetization refrigerator (ADR). \ding{174}: The GGG stage of the ADR. The FAA stage is thermally connected to the cold plate on which the detector array is situated. \ding{175}: The filter system with different colors delineating filters at various temperature stages. The \textit{top-right} panel shows the sizes of the 4 K supporting frame, and the \textit{bottom-right} panel displays the interior structure of the DA.
    }
    \label{fig:struc}
\end{figure}

Fig \ref{fig:struc} shows a preliminary design of the detector assembly (DA) of DIXE. The DA consists principally of the TES microcalorimeter array and the associated SQUID readout electronics, magnetic shields and a set of optical blocking filters. 
The design of the X-ray detectors involves a TES microcalorimeter array with 10$\times$10 pixels at the mK stage, each with a power consumption of about 1 pW. These microcalorimeters are required to have an energy resolution of 6 eV or better over an energy range of 0.1-10 keV. The readout system adopts a frequency-division multiplexing (FDM) scheme. For DIXE, the design calls for a multiplexing factor of 25, so four FDM modules are needed to readout 100 pixels. 
Each module employs a two-stage SQUID system: the first stage at the cold stage with 2 nW power dissipation and the second stage at the 4K stage with a heat load under 200 nW. The detector (and SQUID electronics as well as Inductor-Capacitor (LC) resonators) is surrounded by the magnetic shields, and looks through the blocking filters. Further details on the development of the TES microcalorimeters are described in \cite{Jin_DIXE_LTD} and \cite{sifan_Wang_2022}. In this paper, we'll be focusing on the design of the DA.

The available space for the DA and the adiabatic demagnetization refrigerator (ADR) cooling system is constrained, due to the limited volume of the compartments (up to $720\rm\,mm\times570\,mm\times720\,mm$) in the transporting vehicle (to the CSS), which presents the most severe challenge in the design and optimization of the DA. Other constraints include the total mass of the payload ($<$200 kg) and the electrical power ($<$1000 W). The lifetime of DIXE is projected to be three years.
The aims of optimizing the DA design are: 1) To provide mechanical support for the TES microcalorimeter detector array and multiplexing readout electronics, ensuring that they can survive the launch; 2) To reduce magnetic field near the detector array through effective shielding, minimizing its effects on the detector and associated superconducting readout electronics; and 3) To block radiation at UV/optical/IR wavelengths through the use of filters, reducing the degradation in the energy resolution of the detector caused by shot noises. 
The optimization should balance structural integrity, magnetic shielding, and shot-noise minimization to achieve the desired energy resolution.

\section{Mechanical and thermal design}
\label{sec:mechanical}
The space constraints limit the size and configuration of the sub-4K unit of the payload, which consists of the DA and the ADR cooling stages. Fig \ref{fig:struc} shows a preliminary design.  The sub-4K unit is arranged in an `L'-shaped configuration, and the sizes are shown in the \textit{top right} panel in Fig \ref{fig:struc}. The ADR has a cooling power of $2~\rm \mu W$ at $100~\rm mK$, corresponding to an expected hold time of $9~\rm h$ (for a duty cycle around 90\%). This necessitates a design that minimizes the heat load on the cold stage while maintaining mechanical robustness.

The mechanical structure of the DA is shown in Fig \ref{fig:struc}. The total mass is about 2.4$~\rm kg$. The detector array and the superconducting readout electronics are mounted on the cold plate, which is supported by six Kevlar bundles of {\color{blue}50} mm in length of the support pillar (see Fig.~\ref{fig:struc} {\it right} panel). The mass of the cold plate needs to be minimized to avoid resonance vibration at low frequencies. So the nonuniform thickness of the cold plate is designed to reduce the mass, with thick layers $4~\rm mm$ and thin layers $1~\rm mm$. Six ribs are adopted to stiffen the plate.
The niobium (Nb) shield is connected to and supported by the pillar through interface flanges. The support pillar is screwed onto the flanges, which are made of high-permeability mu-metals for the purpose of the magnetic shield (see \S~\ref{sect:magshield}). The bottom mu-metal flange is mounted on the supporting frame which is cooled by the cryocoolers to 4 K, along with the hybrid magnetic shields, and the support pillars. The cold plate is thermally isolated from the 4 K stage and sits nominally at $50-100$ mK. 

A copper strip is thermally connected with the cold end of the second stage ADR, which cools the cold plate to the transition temperature of TES. The heat load due to thermal conduction through the Kevlar support is expected to dominate other environmental heat input to the cold plate, and the total heat load is required to be less than 1 $\mu\rm W$, in order to stay within the limited cooling power of the ADR. The heat load per Kevlar bundle is given by:
\begin{equation}
    q = \frac{A}{L}\int_{T_{\rm cold}}^{T_{\rm warm}} k(T)\ \mathrm dT,
\end{equation}
where $A$ and $L$ are the cross-section and length of the Kevlar bundle, respectively. Taking the thermal conductivity of Kevlar ($k(T)$) from the literature~\cite{Ventura_2000_kevlar}, we use the equation to compute the heat conducted from the warm end (at $T_{\rm warm}=4\ \rm K$) to the cold end (at $T_{\rm cold}=0.05\ \rm K$), and arrive at a heat load of $\sim0.4\ \rm \mu W$ for six Kevlars with 480 threads per bundle. 

Mechanically, the challenge is for the detector assembly to survive launch vibration. We consider an acceleration level of $\sim 13 \ \rm g$ (rms)~\cite{XQC2002} (although it is expected to be lower for DIXE). For a cold plate of mass $\sim 76\ \rm g$, it corresponds to a launch load of about 9.8 N (rms). We conservatively multiply it by a factor of 8 to arrive at 78 N, which is to be compared with the breaking strength of the Kevlar bundles. For a bundle of 480 threads, it has been shown that the breaking point is at $\sim 163\ \rm N$\footnote{\url{https://www.matweb.com/index.aspx}}, which leaves plenty of margin in the design.


\section{Magnetic shielding}\label{sect:magshield}
\subsection{Requirement}
As superconductor components, the performance of TES microcalorimeter and SQUID readout electronics are susceptible to magnetic fields \cite{Hijmering2013}. The background magnetic field arises mainly from the residual field of the ADR magnet and the Earth field. We have adopted a $\sim4\times10^4~\rm Gauss$ as a starting point for the magnetic shielding design. It is a conservative value as the maximum field wouldn't surpass $4\times10^4~\rm Gauss$ under our ADR operation requirement \cite{Jiang_2023_ADR}.
In this case, the residual magnetic field attenuated by the ADR magnetic shielding is $\lesssim0.1~\rm Gauss$. In contrast,
for the optimal performance of the detectors, the field should be reduced to less than $0.01~\rm Gauss$ \cite{Hijmering2014}. 

\begin{figure}[H]
    \centering
    \includegraphics[width=\textwidth]{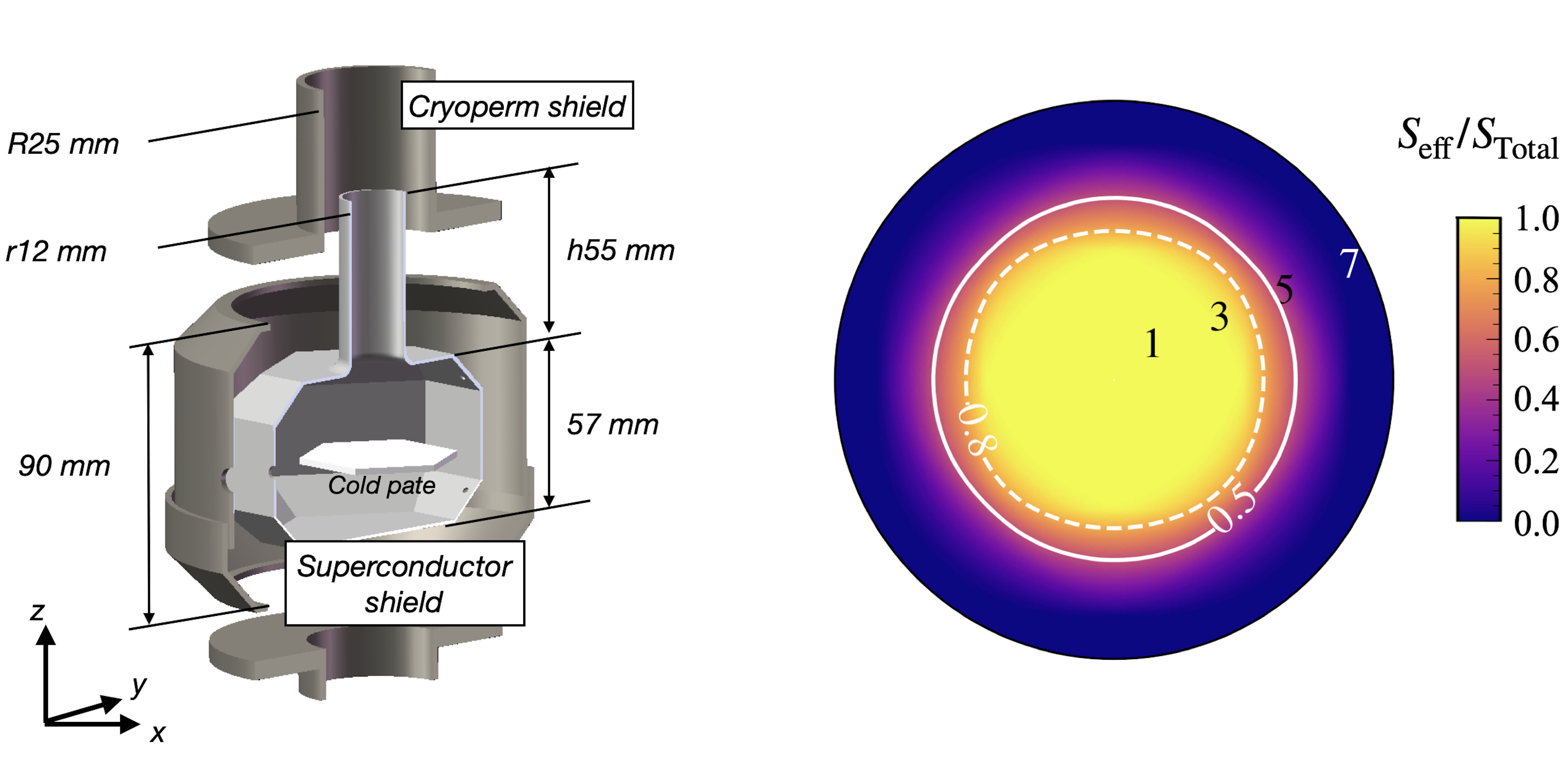}
    \caption{\textit{Left}: Initial design of the magnetic shielding for DIXE. \textit{Right}: Collimator response (see the main text for a definition). The color map in polar coordinates shows the variation of the response over the field of view. The fiducial values of the response and the corresponding off-axis angle are indicated in the map. The dashed and solid curves show the contour with the ratio values of 0.8 and 0.5.}
    \label{fig:shield_design}
\end{figure}

Since the space is limited, multiple configurations of the two-stage ADR and the magnetic shielding system are considered, and a side-by-side design is adopted to minimize the background magnetic field asserted on the magnetic shields (shown in \textit{left} panel of Fig \ref{fig:struc}). The initial design of magnetic shielding is shown in Fig. \ref{fig:shield_design}. It combines a high-permeability outer shield and a superconductor inner shield. Cryoperm, which is used as the material for the outer shield, is a high-permeability Fe/Ni alloy optimized for cryogenic applications\footnote{http://www.cryopermshielding.com/cryoperm-shielding.php}. The high-permeability shield is divided into four parts to enable assembly. The middle two parts have a thickness of $2~\mbox{mm}$, separated by a $1~\mbox{mm}$ gap. The top and bottom parts are thicker ($4-6~\mbox{mm}$) to ensure mechanical robustness, as they also support the pillars (Fig.~\ref{fig:struc}). The inner shield is made of Nb. Note that the bottom of the Nb shield is sealed to enhance magnetic field isolation. As a result, an opening has been made on the right side of both the inner and outer shields. This opening is designed for establishing a thermal connection to the ADR, as detailed and demonstrated in \S~\ref{sec1}. A hexagonal structure is adopted for easier fixation and connection. The thickness of the Nb shield is $1\ \rm mm$. 

In this design, it is the outermost blocking filter (at 300 K) that limits the field of view (FoV) of the payload (see \S~\ref{sec:filter}). We define the collimator response as the ratio of the effective area $S_{\rm eff}$ to the total area of the detector array $S_{\rm TES}$, and compute it as a function of the incident angle ($\theta$) of incoming X-rays. The results are also shown in Fig.~\ref{fig:shield_design}. For a radius of $12~\rm mm$ and a height of 130 mm above the detector plane, the offset angle is about $5^\circ$ as the ratio decreases to 0.5. This leads to a full-width half maximum (FWHM) of $10^\circ$ (shown as the solid white contour on the \textit{right} panel of Fig \ref{fig:shield_design}), which meets the requirement of DIXE. The neck portion of the inner shield must not block the FoV.

\subsection{Static Modeling}
We use COMSOL Multiphysics with the AC/DC module for magnetic shielding simulations. In the static simulation, A zero-field cooling (ZFC) process is assumed, and a constant relative permeability $\mu_r=20000$ for the Cryoperm shield \cite{Bergen_2016_RScl} and ADR shield is adopted. The lower critical magnetic field of Nb at 4 K is $H_{\rm c1}\sim 1444 \ \rm Oe$ ($\sim 18\ \rm Gauss$) \cite{Finnemore_Nb}. We assume a background magnetic field of $B_{\rm res}\lesssim 1 \ \rm Gauss$, and ignore the magnetic flux pinning effect in type II superconductor (Nb). 

We start the modeling by applying a constant magnetic field $\mathbf B_{\rm res}$ along the axial ($z$) and radial $(x, y)$ direction. The $x$ direction goes across the center of the hole for thermal connections. The underlying physics is described by Maxwell's equations:
\begin{equation}
    \begin{aligned}
        & \nabla \times \mathbf{H} = \mathbf{J} \\
        & \mathbf{B} = \nabla\times \mathbf{A} + \mathbf{B}_{\rm res} \\
        & \mathbf{J} = \sigma \mathbf{E} \\
        & \mathbf{B} = \mu_0\mu_r \mathbf{H}
    \end{aligned}
\end{equation}
The effectiveness of the system can be quantified by the \textit{shielding factor}, defined as:
\begin{equation}
    S = \frac{B_{\rm TES}}{B_{\rm res}},
\end{equation}
where $B_{\rm TES}$ is the magnetic field at the position of the TES detector array and $B_{\rm res}$ is the residual background magnetic field.

\begin{figure}[H]
    \centering
    \includegraphics[width=0.5\textwidth]{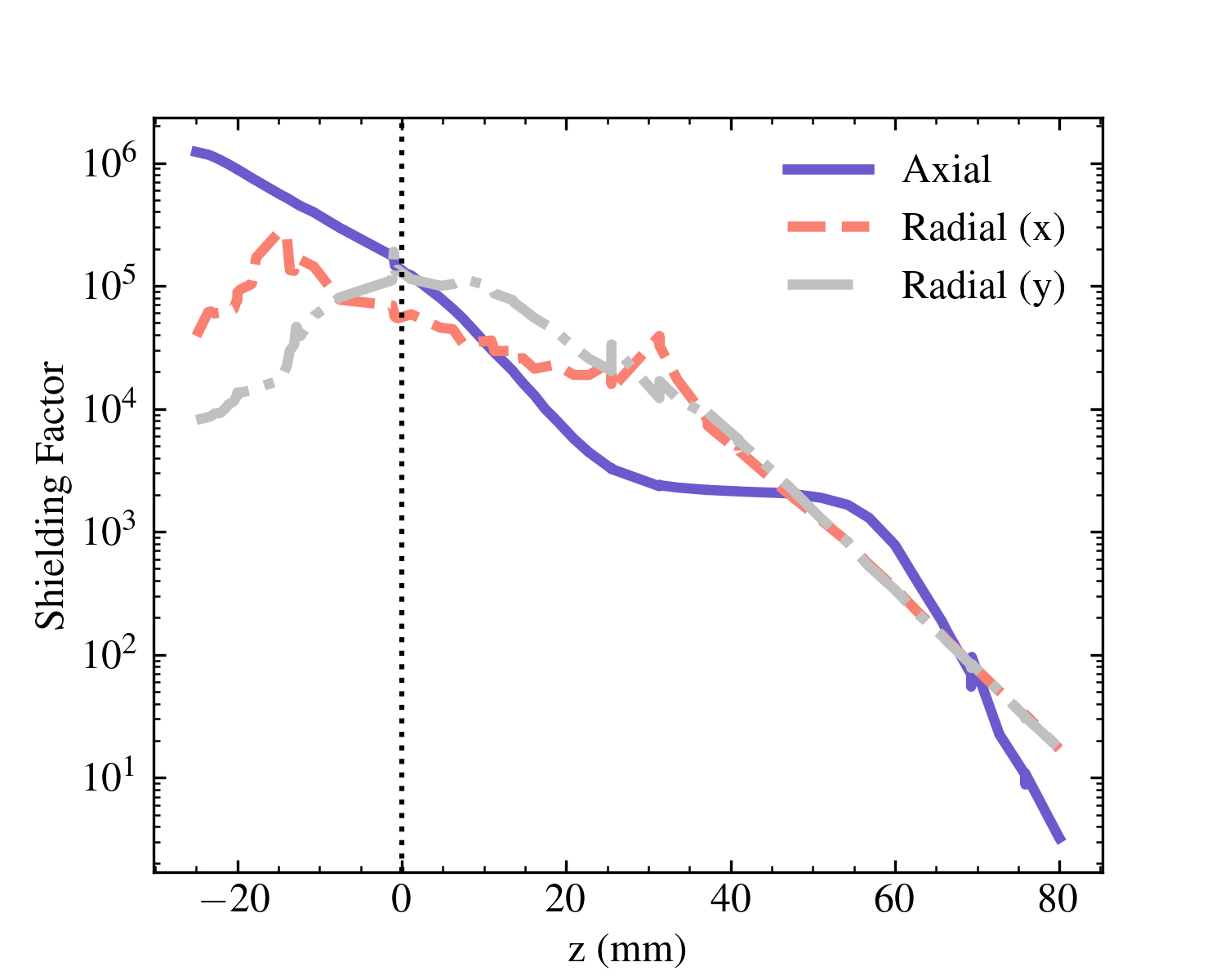}
    \caption{Simulated shielding factors. The solid-blue, dashed-salmon, and dash dotted-gray curves show the variation of shielding factors along the axial $z$, radial $x$, and $y$ directions, respectively, for a constant background field of 1 Gauss. The dotted line indicates the position of the cold plate.}
    \label{fig:constB}
\end{figure}

We define an 80 mm thick sphere as the infinite element domain to model the field boundary conditions at infinity. For the static simulation, we adopt the \textit{Magnetic Insulation} condition for the superconductor shield, forcing $\mathbf{n}\times\mathbf{A}=0$ at the shield boundaries, where $\mathbf{n}$ is the normal vector of the shield's surfaces. The result is shown in Fig \ref{fig:constB}. The shielding factor is above $10^5$ for an axial background field and between $10^4-10^5$ for a radial background field. The magnetic field after attenuation should be less than $1\ \rm\mu T$ with $\sim1\ \rm Gauss$ background field.

\begin{figure}[H]
    \centering
    \includegraphics[width=\textwidth]{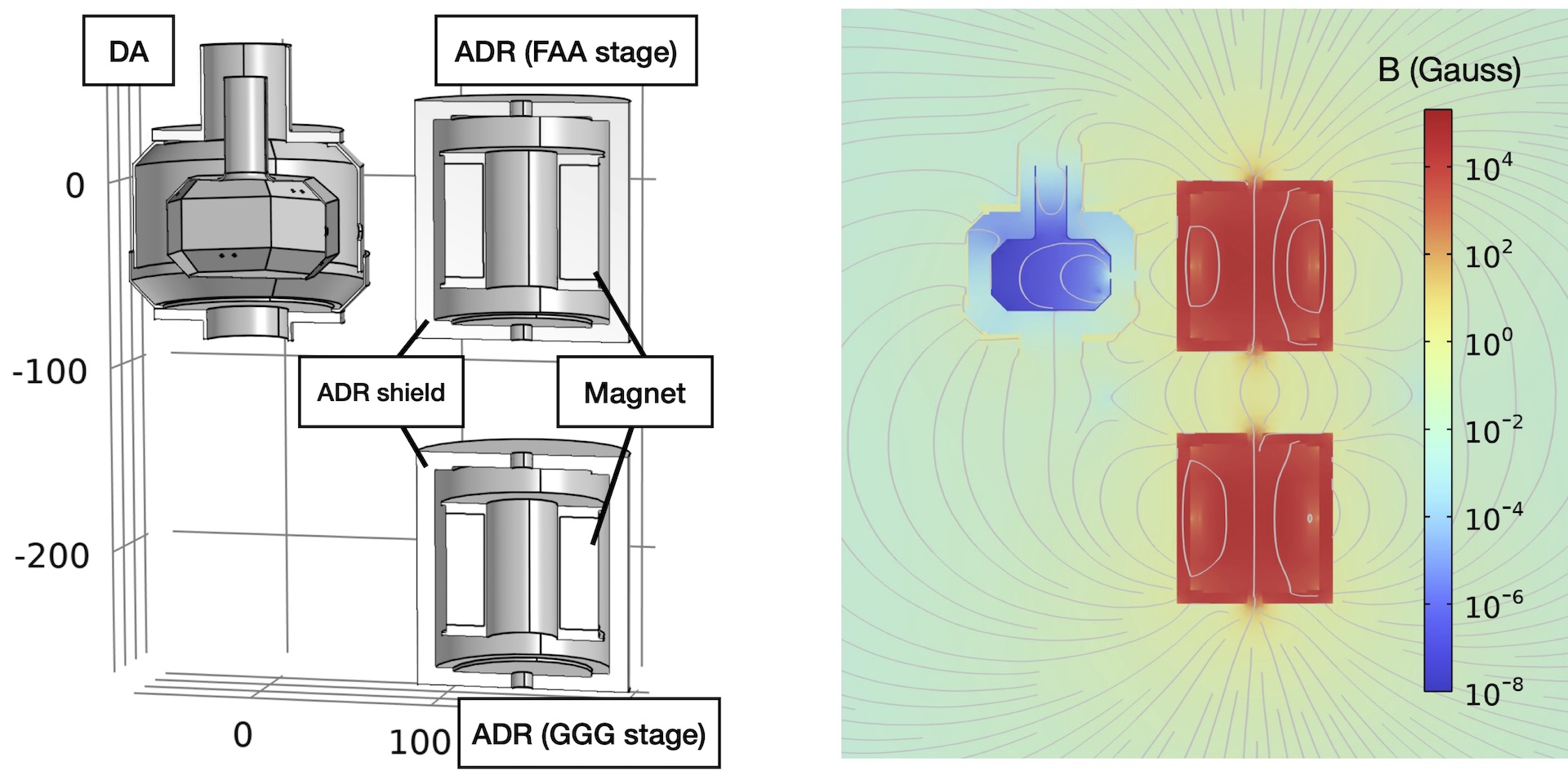}
    \caption{Distribution of the simulated magnetic field based on the initial design of the magnetic shielding. \textit{Left}: Geometric setup for the COMSOL simulation. The magnetic shielding for the detector assembly and for the ADR model are both shown. \textit{Right}: Distribution of the simulated magnetic field. The colored map shows the value of the magnetic field in Gauss with the field lines in grey. }
    \label{fig:magsim}
\end{figure}

In real cases, the residual magnetic field produced by the ADR magnet is complicated. During the operation, the magnetic field produced by the superconducting coil in the ADR is attenuated by another Cryoperm shield surrounding the ADR. Here, we simulate the shielding effects for the full field. The model setup is shown in Fig. \ref{fig:magsim}, and the geometric parameters are summarized in Table \ref{tab:geoparams}. The magnetic field in the superconducting coil is simulated by applying a counterclockwise current (as viewed from the $z$ direction). The current density is assumed to be $J\sim 2\times10^8 \ \rm A/m^2$, which leads to a central field of about $4~\rm T$ and a background field of $< 0.1 \ \rm Gauss$. Taking into account the necessary holes for Kevlar suspensions and the thermal connection, the magnetic field at the position of the detector array is $\sim 10^{-7}-10^{-6}\ \rm Gauss$. For an initial design, we adopt the most conservative approach by adding a 10 mm thick mumetal shield for the ADR. Cryoperm is not utilized for the ADR shield since the magnetic field generated by the magnet surpasses the saturated magnetic field capacity of Cryoperm, which reduces the shielding factor. A thinner mumetal shield for ADR could be considered to reduce the total mass, which would result in a higher background field. Even without a shield around the ADR, the field inside the inner shield would be only $B_{\rm TES}\sim 10^{-4}\ \rm Gauss$ when the magnetic field produced by the superconducting magnet is $\sim4 \ \rm T$.

\begin{table}
    \centering
        \caption{The design parameters (in units of mm) both for the magnetic shielding of the detector assembly and for the ADR. Here, $r_{\rm body}$ refers to the outer radius of the main body of a shield, while $r_{\rm op}$ to the inner radius of the neck of the shield; $d$ indicates the thickness of the main boy of the shield, and $h_{\rm op}$ the height of the neck. 
        }
    \renewcommand{\arraystretch}{1.2}
    \newcolumntype{Y}{>{\centering\arraybackslash}X}
    \begin{tabularx}{\textwidth}{XYYY}
    \hline
    \hline
        Geometric  & Cryoperm shield & Niobium shield & ADR shield \\
        parameters & & & \\
    \hline
        $r_{\rm body}$ & 63 & 50 & 57.5 \\
        $d$ & 2 & 1 & 10 \\      
        $r_{\rm op}$ & 22.5 & 12 & 5 \\
        $h_{\rm op}$ & 45  &  55 &  123 \\
    \hline
    \end{tabularx}
    \label{tab:geoparams}
\end{table}

\section{Filter system}\label{sec:filter}
\begin{figure}[H]
    \centering
    \includegraphics[width=0.6\textwidth]{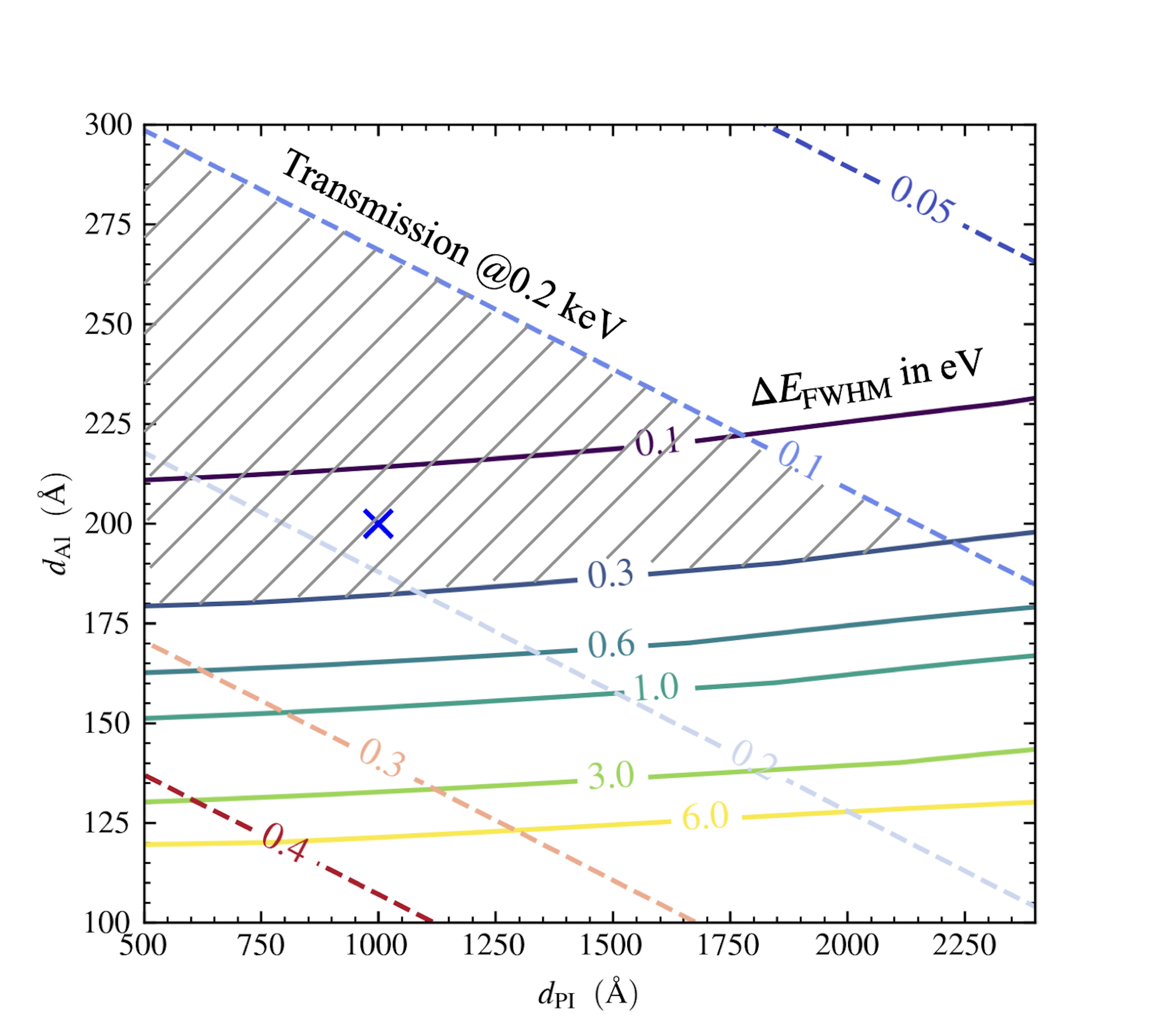}
    \caption{Degradation in energy resolution $\Delta E_{\rm deg}$ (in solid lines) in the unit of eV and X-ray transmission (in dashed lines) as a function of the thickness of the polyimide substrate and of the aluminum layer. The X-ray transmission is computed at an energy of 0.2 keV. The hatched area shows the parameter space that satisfies the DIXE requirements, with the blue cross indicating the values adopted for the preliminary design. 
   }
    \label{fig:d_al_pi}
\end{figure}

Blocking filters are used to minimize the shot noise generated by UV/optical/infrared photons, which degrades the energy resolution of the detectors, as the microcalorimeters also respond to those photons. 
Following the design adopted by the X-ray Quantum Calorimeter (XQC) sounding rocket experiment~\cite{XQC2002}, we examine the filters based on aluminum-coated polyimide (PI).
For an initial design, we place five filters at different temperature stages, as summarized in Table \ref{tab:filter}. The requirement is that the degradation of energy resolution should be less than $5\%$ (i.e., $0.3\ \rm eV$ for a detector resolution of 6 eV). In addition, the absorption by the filters needs to be minimized at the lowest energies, as DIXE is designed to cover an energy range down to $0.1~\rm keV$ (based on its key science objectives~\cite{Jin_DIXE_LTD}).  The requirement is that the transmission exceeds 10\% at 0.2 keV. Fig. \ref{fig:d_al_pi} shows these constraints and the allowed parameter space (shown as the hatched area) for the thickness of the Aluminum and Polyimide films.

We calculate the reflection and transmission coefficients for the filters based on the matrix form of wave propagation theories. The refractive index of the filter material is used in the calculation. Oxidation of aluminum is also considered. The aluminum oxide layer occurs at both sides of the aluminum film with a thickness of $\sim30~\rm\AA$ \cite{XQC_imporvment}. This is because polyimide is transparent to oxygen, exposing the bottom side of the aluminum (the side connected to the PI layer) to oxygen as well. The values of refractive index and extinction coefficient for Al, PI, and $\rm Al_2O_3$ used in the calculation are referred to Hagemann et al. \cite{hagemann1975optical}, Cavidi et al. \cite{Cavadi_2001}, and Querry \& Marvin \cite{querry_1985_optical}.

\begin{figure}
    \centering
    \includegraphics[width=\textwidth]{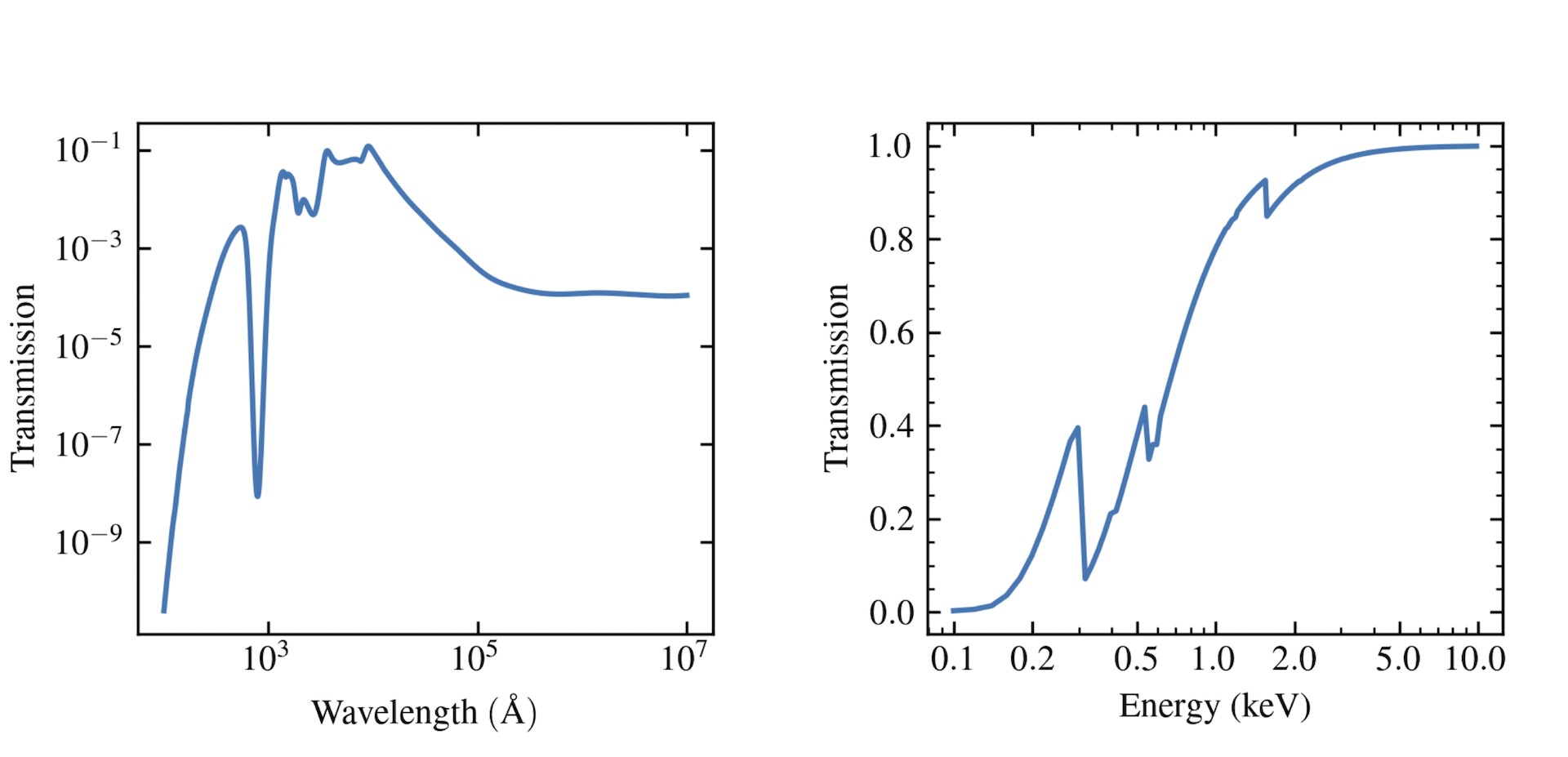}
    \caption{The transmission of UV/Optical/IR photons for a single filter (\textit{left}) and X-rays (\textit{right}) with five filters. Each filter is made of a 200 $\rm \AA$ aluminum film coated on a 1000 $\rm \AA$-thick polyimide film. A $30~\rm\AA$ thick $\rm Al_2O_3$ film on both sides of aluminum is also considered.}
    \label{fig:filter_trans}
\end{figure}

For the calculation of energy resolution degradation $\Delta E_{\rm deg}$, we treat each filer as a blackbody source, and $\Delta E_{\rm deg}$ is given by \cite{Barbera_2015_SPIE}:
\begin{align}
    &{\rm NEP}^2(T)=\frac1{N_{\rm pix}}\int_{\lambda_{\text{min}}}^{\lambda_{\text{max}}}P_\lambda(T) \frac{hc}\lambda\,\mathrm d\lambda, \\
    & \Delta E_{\text{deg}} = 2.35\times 6.2415\times10^{18}\times\sqrt{{\rm NEP^2}(T)\tau},
\end{align}
where ${\rm NEP}$ is the noise equivalent power in the units of $\rm W/\sqrt{Hz}$, $\Delta E_{\rm deg}$ the corresponding energy resolution in the units of eV,
$P_\lambda(T)$ the spatially integrated, wavelength-weighted radiation power arriving at the TES array in the units of $\rm W/\AA$, $N_{\rm pix}$ the number of pixels, and $\tau$ the integration time. Fig. \ref{fig:d_al_pi} shows the degradation in energy resolution due to the shot noise ($\Delta E_{\rm deg}$) and the X-ray transmission at $0.2\ \rm keV$, over a range of the thickness of the aluminum layer and the polyimide substrate. Based on the requirements of the energy resolution degradation and X-ray transmission, the allowed parameter space is limited to the hatched area. Here, we adopt $1000\ \rm\AA$ and $200\ \rm \AA$ for the polyimide and aluminum layers, respectively. This combination is proven to perform well and is less challenging for fabrication \cite{XQC_imporvment}. The transmission curve over a wavelength range covering UV, optical, and IR for a single filter is shown in Fig. \ref{fig:filter_trans} (\textit{left panel}). Also shown in the figure (\textit{right} panel) is the X-ray transmission curve for the case of all five filters. The radiative power from each filter and the contribution to shot-noise-caused energy resolution degradation are summarized in Table \ref{tab:filter}.

\begin{table}
    \centering
    \renewcommand{\arraystretch}{1.2}
    \newcolumntype{Y}{>{\centering\arraybackslash}X}
        \caption{The properties of the five filters and the contribution for radiative heat load and energy resolution degradation. The columns are: (1) The index for each filter; (2) The temperature stage for each filter in the unit of K; (3) The distance between the filter and the TES detector array in the unit of mm; (4) The radiative power contributed by each filter in the unit of W; (5) The energy resolution degradation in the unit of eV.}
    \begin{tabularx}{\textwidth}{YYYYY}
    \hline
    \hline
    Filter index  &  $T$ & $z$ &  $P$  & $\Delta E_{\rm FWHM}$ \\
    (1) & (2) & (3) & (4) & (5)  \\
    \hline
    1 & 1 & 80 & $1.9\times10^{-17}$ & 0.0043 \\
    2 & 4 & 86 & $7.2\times10^{-16}$ & 0.031 \\ 
    3 & 20 & 100 & $8.8\times10^{-17}$ & 0.020 \\
    4 & 77  & 115 & $2.1\times10^{-18}$ & 0.0066 \\
    5 & 300 & 130 & $1.0\times10^{-16}$ & 0.16 \\
    Total &  &  & $9.3\times10^{-16}$ & 0.16\\
    \hline
    \end{tabularx}
    \label{tab:filter}
\end{table}

\section{Discussion}
In the present design, the heat load to the cold plate through the Kevlar support is still a bit high (see \S~\ref{sec1}), compared with the cooling power (estimated to be $\sim2~\mu$W) of the ADR. A reduction by a factor of 2 might be feasible by using fewer Kevlar threads, or by lowering the temperature gradient along the Kevlar support with a thermal link from the 1~K stage to the middle of the Kevlar bundle. 

The design of the ADR, and notably the size of the salt pill, impacts the design of the DA magnetic shielding by dictating the necessary maximum magnetic field. For our ADR, with a target hold time of 9 hours and a duty cycle of 90\%, an increase in the salt pill's mass reduces the maximum magnetic field. This reduction allows for smaller and more compact DA magnetic shields. The trade-off, however, is a considerable increase in the total mass of the salt pill-magnet assembly, as well as adversely affecting the ADR's duty cycle. Future studies will aim to refine the ADR design further.


Not yet considered is the need for anticoincidence detectors. Geant~4 simulations are being carried out to assess the particle background.

\section{Conclusion}
A preliminary design of the DIXE DA is presented and analyzed. It meets the requirements of mechanical robustness, magnetic shielding, heat load, and energy resolution degradation.

\bmhead{Acknowledgments}

We wish to thank Dr. Dan McCammon and all members of the DIXE collaboration team for useful discussion. This work was supported in part by the Ministry of Science and Technology of China through Grant 2022YFC2205100, by China National Space Administration (CNSA) through a technology development grant, and by the National Natural Science Foundation of China through Grants 11927805, 12203027, 11803014, and 12220101004.

\bibliography{main}

\end{document}